\documentclass[aps, prd, amsmath, floats, floatfix, twocolumn, nofootinbib]{revtex4-1}
\pdfoutput=1
\usepackage{graphicx}
\usepackage{color}
\usepackage{latexsym}

\usepackage{amssymb}

\newcommand{\be}{\begin{eqnarray}}
\newcommand{\ee}{\end{eqnarray}}

\newcommand{\bea}{\begin{eqnarray}}
\newcommand{\eea}{\end{eqnarray}}

\begin{document}

\title{The Higgs mechanism in nonlocal field theory}

\author{Leonardo Modesto}
\email{lmodesto@sustech.edu.cn}

\affiliation{Department of Physics, Southern University of Science
and Technology, Shenzhen 518055, China}

\begin{abstract}
We provide an example of nonlocal scalar electrodynamics that allows the same Higgs mechanism so successful  in local field theory. The nonlocal action is structured in order to have the same exact solutions and the same equations of motion for perturbations of the local theory, at any perturbative order. Therefore, the perturbative degrees of freedom that propagate in the unstable vacuum are reshuffled when the stable vacuum is replaced in the EoM, but their number does not change at any perturbative order, and their properties are the same like in the usual local theory. Finally, the theory is super-renormalizable or finite at quantum level.

\end{abstract}

\maketitle


In two quite recent interesting works \cite{Gama:2018cda,Hashi:2018kag} has been addressed the issue of the Higgs mechanism in nonlocal field theory. In both such papers, it has been pointed out that the number of degrees of freedom is not preserved when the usually stable non tachyonic vacuum is replaced in the action. Indeed, after spontaneous symmetry breaking we can have more real degrees of freedom including an infinite number of complex conjugate poles in the propagator \cite{shapiro3,Shapiro:2015uxa,ModestoShapiroLW, ModestoLW}, which signal vacuum instabilities. 
In \cite{Hashi:2018kag}, the authors proposed a purely nonlocal gauge theory in six dimensions in order to avoid the proliferation of extra degrees of freedom. 
In this paper we advance a nonlocal formulation of the scalar quantum electrodynamics 
based on four requirements that turned out to be crucial in order to define stable nonlocal gravitational theories \cite{Krasnikov,kuzmin,modesto,modestoLeslaw,review} in presence or not of matter \cite{Universally,FiniteGaugeTheory} and with or without supersymmetry \cite{Giaccari:2016kzy,Calcagni:2014vxa}.
The required properties are:
 (i) all the exact solutions of the local theory must be solutions of the nonlocal theory too \cite{Li:2015bqa}, (ii) the equations of motion (EoM) for the perturbations around the latter background solutions must be the same of the local theory, (iii) the local and nonlocal theory must have the same tree-level scattering amplitudes (this guarantees macro-causality \cite{scattering,nonlocaldesitter,causality}), 
(iv) the nonlocal field theory is unitary \cite{cutkosky,Briscese:2018oyx, Briscese:2021mob, PiusSen} and super-renormalizable or finite at quantum level 
\cite{Krasnikov,kuzmin,modesto,modestoLeslaw,review}.  




The theory is defined in terms of the following five quantities: a nonlocal action $S[\Phi_i]$, a local Lagrangian with $U(1)$ gauge symmetry $\mathcal{L}$, the local EoM $E_i$, a nonlocal form factor $F(\Delta)$, whose argument $\Delta$, which is the Hessian of the local theory,  
\be
&& S[\Phi_i] = \int {\rm d}^D x \left( {\mathcal L} + E_i \mathcal{F}_{ij} E_j  
\right) \, , 
\label{actionFH} \\
&& \mathcal{L} = - \frac{1}{4} F_{\mu\nu} F^{\mu\nu} 
 + (\mathcal{D}_\mu \phi )^*  (\mathcal{D}_\mu \phi ) - V(\phi)
 \, ,
 \label{localL}
 \\
&& E_i(x) =  \frac{\delta  S_\ell}{\delta \Phi_i(x)} \, , \quad S_\ell = \int d^D x \mathcal{L} \, , \label{LocalEoM} 
\\
 &&
2 \Delta_{ik} \mathcal{F}_{k j} \equiv \left( e^{H(\Delta_\Lambda)} - 1\right)_{i j} \, , 
 \label{Effe}
 \\
 && \Delta_{i j}(y, x)  \equiv  \frac{ \delta E_j(x)}{\delta \Phi_i(y) 
 } \, ,
\label{Delta} 
\ee
where we defined $\Phi_i = (A_{\mu}, \phi, \phi^*)$, $H$ is an entire analytic function of its argument, and h.c. stays to the hermitian conjugate of the second operator in the action quadratic in the local EoM. 
The reader can explicitly verify that $\Delta_{ij}(x,y)$ = $\Delta_{ji}(y,x)$ when such operator is integrated in the variables $x$ and $y$ \cite{NLSTM}.

The potential $V(\phi)$ and the gauge covariant derivative are:
\be
&& V(\phi) = - \mu^2 \phi^*\phi  + \lambda (\phi^*\phi)^2 \, , 
\label{potential}\\
&& \mathcal{D}_{\mu} = \partial_\mu + i e \, A_\mu \, , 
\ee
with $\mu^2 > 0$. 
We can also decompose the complex field $\phi$ in a real and an imaginary part in order to express the Lagrangian in terms of two real scalar fields, namely
\be
\phi = \frac{1}{\sqrt{2}} \left( \phi_1 + i \phi_2 \right) \, .
\label{phi12}
\ee
The potential (\ref{potential}) has a maximum in $\phi = 0$, while the minimum is located in
\be
\phi^* \phi = \frac{1}{2} \left(\phi_1^2 + \phi_2^2 \right) = \frac{\mu^2}{2 \lambda} \equiv \frac{\phi_0^2}{2}  \, .
\label{min12}
\ee

Taking the variation of the local Lagrangian (\ref{localL}) respect to the gauge boson $A_\mu$ and the two scalar fields $\phi$ and $\phi^*$ the local EoM (\ref{LocalEoM}) read:
\be
&& E_A = \partial_\mu F^{\mu\nu} - i e \, \left[ \phi^* (\mathcal{D}^{\nu} \phi ) - (\mathcal{D}^{\nu} \phi )^* \phi \right] \, , 
\label{AEoM} \\
&& E_\phi = - \mathcal{D}^{\mu} (\mathcal{D}_{\mu} \phi )^* - \frac{\partial V}{\partial \phi} \, , 
\label{phiEoM} \\
&& E_{\phi^*} = - \mathcal{D}^*_{\mu} (\mathcal{D}^{\mu} \phi ) - \frac{\partial V}{\partial \phi^*} \, .
\label{phisEoM} 
\ee
Therefore, the EoM for the nonlocal field theory defined by the action (\ref{actionFH}), up to terms quadratic in the local EoM, are: 
\be
{\mathcal E}_k &=&  E_k +  2 
\Delta_{ik} \, 
 \mathcal{F}_{k j} E_j + [ O(E^2) ]_k \nonumber \\
&=&\big( e^{H( \Delta_{\Lambda}) } \big)_{k j } \,  E_j + [ O(E^2) ]_k  = 0 \, .
\label{NLEoM}
\ee
We can invert the exponential in (\ref{NLEoM}) to end up with the following nonlocal EoM, 
\be
\mathcal{\tilde{E}}_i \equiv E_i + \big( e^{-H( \Delta_{\Lambda} ) } \big)_{i j } \left[ O(E^2) \right]_j = 0 \, .
\label{tildeEoM}
\ee
Since the operator $\exp {-H( \Delta_{\Lambda} ) }$ is invertible, we have:
\be
\mathcal{\tilde{E}}_i \equiv \left( e^{-H( \Delta_{\Lambda} ) } \right)_{i j} \mathcal{{E}}_j= 0 \quad  \Longleftrightarrow \quad  \mathcal{E}_i = 0 \, , 
\label{original}
\ee

%
Now, given an exact solution of the nonlocal EoM ($\mathcal{\tilde{E}}_i =0$) compatible with $E_i =0$, we can derive the EoM for the perturbations defined through an expansion of the fields, and of the local and nonlocal EoM, in a small dimensionless parameter $\epsilon$, namely 
\be
&&\hspace{-0.5cm} 
 \Phi_i = \sum_{n=0}^{\infty} \epsilon^n \Phi_i^{(n)} \, , \\
&& \hspace{-0.5cm} 
E_k(\Phi_i) =  \sum_{n=0}^{\infty} \epsilon^n E_k^{(n)} \, , \quad 
 \mathcal{\tilde{E}}_k(\Phi_i) =  \sum_{n=0}^{\infty} \epsilon^n \mathcal{\tilde{E}}_k^{(n)} . 
 \label{expansion}
\ee
The fields $\Phi_i^{(0)}$ satisfy the background EoM, namely 
\be
E^{(0)}_k( \Phi_i^{(0)} ) =0 \,  .
\ee
Hence, it is simple to prove the following theorem, which is a slight generalization of the theorems proved in \cite{StabilityMinkAO, StabilityRicciAO}. 

{\bf Theorem}. In the nonlocal theory (\ref{actionFH}), all perturbations satisfy the same EoM of the perturbations of the local theory defined by (\ref{localL}), namely 
\be
\hspace{-0.6cm}
{\mathcal{ \tilde{E}}^{(n)}_k}(\Phi_i^{(n)}) = 0 \quad \Longrightarrow \quad { E^{(n)}_k} (\Phi_i^{(n)}) = 0 \,\,\,\, \mbox{for} \,\,\,\, n>0 \, ,
\label{LeoTheorem}
\ee
where the label ``$n$" stays for the perturbative expansion of the nonlocal $\mathcal{\tilde{E}}_k$ and local EoM ${E_k}$ at the order ``$n$" in all the perturbations $\Phi_i^{(n)}$. 
Finally, 
\be
\hspace{-0.6cm}
{\mathcal{ {E}}^{(n)}_k}(\Phi_i^{(n)}) = 0 \quad \Longrightarrow \quad { E^{(n)}_k} (\Phi_i^{(n)}) = 0 \,\,\,\, \mbox{for} \,\,\,\, n>0 \, ,
\label{LeoTheorem2}
\ee
because of the double implication (\ref{original}).
Notice that at any perturbative order the exponential form factor always appears diagonal and at the zero order in the perturbations. 

The proof is a straightforward consequence of the EoM (\ref{tildeEoM}), which coincide with the local EoM (\ref{AEoM}), (\ref{phiEoM}), (\ref{phisEoM}), namely $E_k =0$, up to operators $O(E^2)$, and of the invertibility of the exponential form factor in (\ref{NLEoM}) (for more details see \cite{StabilityMinkAO, StabilityRicciAO, NLSTM}). 

We now apply the theorem above at the linear perturbative order to investigate the spectrum of the theory around the vacuum solutions: 
\be
&& 1) \quad \phi = 0 \, , \quad A_\mu = 0 \, , 
 \label{tachyon} 
\\
&& 2) \quad \phi = \phi_0 = \sqrt{ \frac{\mu^2}{\lambda} }
\, , \quad A_\mu = 0 \, , 
 \, ,
\label{nonT}
\ee
%
Since all the solutions of the local theory are solutions of the nonlocal too, one can easily see that the spectrum 
around the tachyonic vacuum (\ref{tachyon}) consists on a massless vector field and two real scalar fields. Indeed, all the other operators in (\ref{tildeEoM}) do not contribute at the linear level 
because quadratic in the local EoM, namely $O({\bf E}^2)$. 

Similarly, when we replace the vacuum (\ref{nonT}) in the nonlocal EoM (\ref{tildeEoM}) by means of the following definition, 
\be
\phi = \frac{1}{\sqrt{2}} ( \phi_0 + \phi_1 + i \phi_2) \, , 
\label{nonTd}
\ee
we get the same spectrum of the local theory with the $U(1)$ symmetry spontaneously broken. Let us expand on this statement reminding the Higgs mechanism in the EoM formulation instead of using the Lagrangian formalism. This derivation is of course equivalent, but as far as we know usually not presented in the textbooks. 

When we replace (\ref{nonTd}) in the EoM (\ref{AEoM})-(\ref{phisEoM}) and we choose the unitary gauge, namely $\phi_2 =0$ (or equivalently $\phi = \phi^*$), we get the following EoM, 
\be
&& E_A = \partial_\mu F^{\mu\nu} +  e^2 \phi_0^2 A^\nu + \dots = 0 \, , 
\label{AEoMv} \\
&& E_\phi = - (\Box + 2 \mu^2 ) \frac{\phi_1}{\sqrt{2}}  + \dots = 0 \, , 
\label{phiEoMv} \\
&& E_{\phi^*} =   (E_\phi)^* \, , 
\label{phisEoMv} 
\ee
where the dots stay for interaction terms at least quadratic in the fields. One of the terms in (\ref{phiEoMv}) is complex and proportional to $i e \phi_1 (\partial_\mu A^\mu)$ (similarly: one of the terms in (\ref{phisEoMv}) is complex and proportional to $ - i e \phi_1 (\partial_\mu A^\mu)$), hence, taking the difference $E_{\phi^*} -   E_\phi^*=0$ one of the two EoM can be replaced with $\partial_\mu A^\mu$, which is the on-shell transversality 
condition for the massive vector field. 
The latter results can now be directly exported to the EoM (\ref{tildeEoM}) 
because the nonlocal EoM match the local ones up to interaction terms quadratic in ${\bf E}$. 

We now investigate the spectrum of the theory expanding the nonlocal action at the second order in the fields. 
The purpose of this paragraph is twofold, on the one hand we will verify the correctness of the newly derived spectrum, on the other hand we will have the nonlocal action in the suitable form for the derivation of the propagators
 and infer about the renormalizability.
 
 \begin{widetext}
In terms of the fields defined in (\ref{phi12}) the local Lagrangian reads:
\be
\mathcal{L} & = & - \frac{1}{4} F_{\mu\nu} F^{\mu\nu} + \mathcal{L}(\phi_1, \phi_2) \, , \\
\mathcal{L}(\phi_1, \phi_2) \! &  =  & \! \frac{1}{2} (\partial_\mu \phi_1)^2 +  \frac{1}{2} (\partial_\mu \phi_1)^2  
-  e \phi_2 (\partial_\mu \phi_1) A^\mu  
 +  e \phi_1 (\partial_\mu \phi_2) A^\mu + \frac{e^2}{2} \left(\phi_1^2 + \phi_2^2 \right) A_\mu A^\mu 
- \mu^2 \left(\phi_1^2 + \phi_2^2 \right) + \lambda \left(\phi_1^2 + \phi_2^2 \right)^2   \!\!  .
\nonumber 
\ee
Given the minimum of the potential in (\ref{min12}) 
the $U(1)$ symmetry is spontaneously broken whether we select the vacuum ($\phi_1 = \phi_0, \,\phi_2 =0, \, A_\mu=0$).

The functional Taylor expansion of the nonlocal action at the second order in the fields $\Phi_i$ can be derived from (\ref{actionFH}), and in short notation (see \cite{NLSTM}) it reads:
\be
S^{(2)} &=& \frac{1}{2} \int d^D x \, \delta \Phi_i  \left[ \frac{\delta^2 S_\ell  }{\delta \Phi_i \delta \Phi_j} + 2 \frac{\delta E_k}{\delta \Phi_i} \mathcal{F} _{k l} \frac{\delta E_l}{\delta \Phi_j}
\right]  \delta \Phi_j 
= \frac{1}{2} \int d^D x \, \delta \Phi_i  \frac{\delta E_k}{\delta \Phi_i} \left[ \delta_{k j} + 2  \mathcal{F}_{k l} \frac{\delta E_l}{\delta \Phi_j}
\right]  \delta \Phi_j \nonumber \\
& = &
\frac{1}{2} \int d^D x \, \delta \Phi_i  \frac{\delta E_k}{\delta \Phi_i} \left[ \delta_{k j} + 2  \mathcal{F}_{k l} 
\Delta_{j l} 
\right]  \delta \Phi_j 
=
\frac{1}{2} \int d^D x \, \delta \Phi_i  \frac{\delta E_k}{\delta \Phi_i} \left[ \delta_{k j} + 2  \mathcal{F}_{k l} 
\Delta_{l j } 
\right]  \delta \Phi_j \nonumber \\
& = &
\frac{1}{2} \int d^D x \, \delta \Phi_i  \frac{\delta E_k}{\delta \Phi_i} \left[ \delta_{k j} + 
\left( e^{H(\Delta)} - 1\right)_{k j} \right]  \delta \Phi_j 
=
\frac{1}{2} \int d^D x \, \delta \Phi_i  \frac{\delta^2 S_\ell  }{\delta \Phi_i \delta \Phi_k}
 \left( e^{H(\Delta)}  \right)_{k j} \delta \Phi_j  \nonumber \\
 & = &
\frac{1}{2} \int d^D x \, \delta \Phi_i  \, \Delta_{i k}
 \left( e^{H(\Delta)}  \right)_{k j} \delta \Phi_j \, ,
 \label{ActionHessian}
\ee
where $\delta \Phi_i = (\delta \phi_1 = \phi_1 - \phi_0 \equiv \varphi, \phi_2 \, , \, A_{\mu} )$ and we used that the Hessian is symmetric under integration.   

Now, in oder to get the second order expansion of the action in the fields $\varphi, \phi_2, A_\mu$ it is convenient to make the following field redefinition,
\be
\delta \Phi_i = \left( e^{ - \frac{H(\Delta)}{2} } \right)_{i j} \delta \tilde{\Phi}_j \, ,
\label{FieldRed}
\ee
which is allowed because the form factor $e^{H(\Delta) }$ is invertible tough $\Delta$ is in general not invertible because of the gauge invariance. Therefore, (\ref{ActionHessian}) turns into:
\be
S^{(2)} & = &
\frac{1}{2} \int d^D x \left[\left( e^{ - \frac{H(\Delta)}{2} } \right)_{i l} \delta \tilde{\Phi}_l\right]   \Delta_{i k}
 \left( e^{H(\Delta)}  \right)_{k j} \left( e^{ - \frac{H(\Delta)}{2} } \right)_{j m} \delta \tilde{\Phi}_m   \nonumber \\
 & = &
 \frac{1}{2}
  \int d^D x \left[\left( e^{ - \frac{H(\Delta)}{2} } \right)_{i l} \delta \tilde{\Phi}_l\right]   \Delta_{i k}
 \left( e^{\frac{H(\Delta)}{2}}  \right)_{k j}  \delta \tilde{\Phi}_j   \nonumber \\
 & = &
 \frac{1}{2}
  \int d^D x \int d^Dy \int d^D z \left[\left( e^{ - \frac{H(\Delta)}{2} } \right)_{i l}(x,y) \, \delta \tilde{\Phi}_l(y)\right]   \Delta_{i k}(x,z)
\left[ \left( e^{\frac{H(\Delta)}{2}}  \right)_{k j}  \, \delta \tilde{\Phi}_j  \right](z) \, ,
\label{HessianFR}
 \ee
 where in the last step we explicitly introduced the kernel representation for the first form factor, which is a function of $\Delta$, and for the operator $\Delta$ itself. 

In order to move the first form factor in (\ref{HessianFR}) from the left to the right of $\delta \tilde{\Phi}_l(y)$, we 
consider only one $\Delta$-term of the Taylor's expansion for $\exp  - H(\Delta)/2$, and we introduce the following short notation, 
\be
\int d^D x \int d^D y \, \left[ \Delta_{i l}(x , y) \, \delta \tilde{\Phi}_l(y) \right] V_i(x) \, , \quad V_i(x) \equiv \int d^D z \, \Delta_{i k}(x,z)
\left[ \left( e^{\frac{H(\Delta)}{2}}  \right)_{k j}  \, \delta \tilde{\Phi}_j  \right](z) . \label{HessianFR2}
\ee
Now we can commute $\Delta_{i l}(x , y)$ and $\delta \tilde{\Phi}_l(y)$ because in the kernel representation all the derivatives, which are present in $\Delta$, do not act on the right side but on internal fields, if there are, or the Dirac's distribution on on the right side \cite{NLSTM}.  
Hence, (\ref{HessianFR2}) is equivalent to:
\be
\int \! d^D x \! \int  \! d^D y \,   \delta \tilde{\Phi}_l(y) \, \Delta_{i l}(x , y)   V_i(x) \, , 
  \label{HessianFR3-1}
\ee
where we also removed the square brackets because of the same reason. 

Finally, we make use of the symmetry property of the Hessian under the integral, namely 
\be
\hspace{-0.7cm} 
\int \! d^D x \! \int  \! d^D y \,  \delta \tilde{\Phi}_l(y) \, \Delta_{i l}(x , y)  \,  V_i(x) 
= \int \! d^D x \! \int  \! d^D y \,   V_i(x) \,  \Delta_{i l}(x , y)  \, \delta \tilde{\Phi}_l(y)
= \int \! d^D x \! \int  \! d^D y \,      \delta \tilde{\Phi}_l(y)  \,  \Delta_{ l i }(y, x)  \, V_i(x) 
\label{SymHess}
\, .
\ee
The generalization of the result (\ref{SymHess}) to the form factor in (\ref{HessianFR}) is straightforward, namely 
\be
S^{(2)} & = &
 \frac{1}{2}
  \int d^D x \int d^Dy \int d^D z  \, \delta \tilde{\Phi}_l(y) \, 
  \left( e^{ - \frac{H(\Delta)}{2} } \right)_{ l i }(y,x) \,   \Delta_{i k}(x,z)
\left[  \left( e^{\frac{H(\Delta)}{2}}  \right)_{k j} \delta \tilde{\Phi}_j\right](z) \, .
\label{HessianFR3}
 \ee
Since $\exp - H(\Delta)/2$ commutes with $\Delta$ we end up with
\be
S^{(2)} & = &
 \frac{1}{2}
  \int d^D x \int d^Dy \int d^D z  \int d^Dw \, 
  \delta \tilde{\Phi}_l(y) \, 
  \left( e^{ - \frac{H(\Delta)}{2} } \right)_{ l i }(y,x) 
  \,   \Delta_{i k}(x,z) \,
 \left( e^{\frac{H(\Delta)}{2}}  \right)_{k j} (z,w) \, \delta \tilde{\Phi}_j(w) \nonumber \\
 & = & 
  \frac{1}{2}
  \int d^D x \int d^Dy \int d^D z  \int d^Dw \, 
  \delta \tilde{\Phi}_l(y) \, 
  \,   \Delta_{l i }(y, x) \,
  \left( e^{ - \frac{H(\Delta)}{2} } \right)_{ i k  }(x,z) 
 \left( e^{\frac{H(\Delta)}{2}}  \right)_{k j} (z,w) \, \delta \tilde{\Phi}_j(w) \nonumber \\
 & = & 
  \frac{1}{2}
  \int d^D x 
 \int d^D y  \, 
  \delta \tilde{\Phi}_l(y) \, 
  \,   \Delta_{i k}(y, x) \,
\, \delta \tilde{\Phi}_j(x) \nonumber\\
& = & 
  \frac{1}{2}
  \int d^D x \, 
  \delta \tilde{\Phi}_l \, 
  \,   \Delta_{i k} \, 
\, \delta \tilde{\Phi}_j
  \, ,
\label{HessianFR4}
 \ee
which coincides with the second order variation of the local theory upon the field redefinition (\ref{FieldRed}). Therefore, the local and nonlocal theories have the same perturbative spectrum when the action is expanded in perturbative fluctuations around a vacuum exact solution of the local EoM.  
Notice that the Hessian has to be expanded on the selected vacuum too. 


\end{widetext}

{\it Conclusions ---} We have introduced a recipe to construct general gauge invariant actions consistent with the Higgs mechanism. Indeed, in nonlocal theories the spectrum is very sensitive to the selected vacuum and it is not guaranteed that the number of degrees of freedom will be preserved following spontaneous symmetry breaking. However, the theory propose in \cite{NLSTM}, and applied here to the scalar electrodynamics, provides a very simple and general solution to the issue pointed out in \cite{Gama:2018cda} and \cite{Hashi:2018kag}. 
Finally, the theory is super-renormalizable (or finite whether we introduce several other local operators that do not spoil all the required properties) as proved in \cite{Universally}. 
The generalization to non-abelian Yang-Mills gauge theories is straightforward according with the theory \cite{NLSTM}.

\acknowledgments
In memory of my mother.

\end{document}